\documentstyle[12pt,prd,aps,epsfig,preprint]{revtex}

\begin{document}
\title{Bound energy for the exponential-cosine-screened Coulomb potential }
\author{Sameer M. Ikhdair\thanks{%
sameer@neu.edu.tr} and \ Ramazan Sever\thanks{%
sever@metu.edu.tr}}
\address{$^{\ast }$Department of Physics, \ Near East University, Nicosia, North
Cyprus, Mersin-10, Turkey\\
$^{\dagger }$Department of Physics, Middle East Technical University, 06531
Ankara, Turkey.}
\date{\today
}
\maketitle

\begin{abstract}
An alternative approximation scheme has been used in solving the
Schr\"{o}dinger equation for the exponential-cosine-screened Coulomb
potential. The bound state energ\i es for various eigenstates and the
corresponding wave functions are obtained analytically up to the second
perturbation term.


PACS\ NO: 03.65.Ge
\end{abstract}


\section{Introduction}

\noindent The generalized exponential-cosine-screened Coulomb (GECSC)
potential or the generalized cosine Yukawa (GCY) potential:
\begin{equation}
V\left( r\right) =-\left( \frac{A}{r}\right) \exp (-\delta r)\cos (g\delta
r),
\end{equation}
where $A$ is the strength coupling constant and $\delta $ is the screening
parameter, is known to describe adequately the effective interaction in
many-body enviroment of a variety of fields such as atomic, nuclear,
solid-state, plasma physics and quantum field theory [1,2]. It is also used
in describing the potential between an ionized impurity and an electron in a
metal [3,4] or a semiconductor \ [5] and the electron-positron interaction
in a positronium atom in a solid [6]. The potential in (1) with $g=1$ is
known as a cosine-screened Coulomb potential. The static screened Coulomb
(SSC) potential ($g=0$ case) is well represented by Yukawa form: $%
V(r)=-(\alpha Ze^{2})\exp (-\delta r)/r$ which emerges as a special case of
the ECSC potential in (1) with $A=\alpha Ze^{2},$ where $\alpha
=(137.037)^{-1}$ is the fine-structure constant and $Z$ is the atomic
number, is often used for the description of the energy levels of light to
heavy neutral atoms [7]. It is known that SSC potential yields reasonable
results only for the innermost states when $Z$ is large. However, for the
outermost and middle atomic states, it gives rather poor results. Although
the bound state energies for the SSC potential with $Z=1$ have been studied
[7].

The Schr\"{o}dinger equation for such a potential does not admit exact
solutions. So various approximate methods [8] both numerical and analytical
have been developed \ Hence, the bound-state energies of the ECSC potential
were first calculated for the $1s$ state using numerical [3,8,9] and
analytical [10,11] methods and for the $s$ states by a variational method
[12]. Additionally, the energy eigenvalues of the ECSC potential [13] have
been recalculated for the $1s$ state with the use of the ground-state
logarithmic perturbation theory [14,15] and the Pad\'{e} approximant method.
The problem of determining the critical screening parameter $\delta _{c}$
for the $s$ states was also studied [16].

It has also been shown that the problem of screened Coulomb potentials can
be solved to a very high accuracy [17] by using the hypervirial relations
[18,19,20] and the Pad\'{e} approximant method. The bound-state energies of
the ECSC potential for all eigenstates were accurately determined within the
framework of the hypervirial Pad\'{e} scheme [21]. Further, the large-N
expansion method of Mlodinow and Shatz [22] was also applied to obtain the
energies of the ground and first excited $s-$ states and the corresponding
wave functions. Recently, we studied the bound-states of the ECSC potential
for all states using the shifted large $N-$ expansion technique [23].

In this paper, we investigate the bound-state properties of ECSC potential
using a new perturbative formalism [24] which has been claimed to be very
powerful for solving the Schr\"{o}dinger equation to obtain the bound-state
energies as well as the wave functions in Yukawa or SSC potential problem
[24,25] in both bound and continuum regions. This novel treatment is based
on the decomposition of the radial Schr\"{o}dinger equation into two pieces
having an exactly solvable part with an addi\i tional piece leading to
either a closed analytical solution or approximate treatment depending on
the nature of the perturbed potential.

The contents of this paper is as follows. In section \ref{TM} we breifly
outline the method with all necessary formulae to perform the current
calculations. In section \ref{A} we apply the approach to the
Schr\"{o}dinger equation with the ECSC potential and present the results
obtained analytically and numerically for the bound-state energy values.
Finally, in section \ref{CR} we give our concluding remarks.

\section{The Method}

\label{TM}For the consideration of spherically symmetric potentials, the
corresponding Schr\"{o}dinger equation, in the bound state domain, for the
radial wave function reads

\begin{equation}
\frac{\hbar ^{2}}{2m}\frac{\psi _{n}^{\prime \prime }(r)}{\psi _{n}\left(
r\right) }=V(r)-E_{n},
\end{equation}
with

\begin{equation}
V\left( r\right) =\left[ V_{0}(r)+\frac{\hbar ^{2}}{2m}\frac{\ell (\ell +1)}{%
r^{2}}\right] +\Delta V(r),
\end{equation}
where $\Delta V(r)$ is a perturbing potential and $\psi _{n}(r)=\chi
_{n}(r)u_{n}(r)$ is the full radial wave function, in which $\chi _{n}(r)$
is the known normalized eigenfunction of the unperturbed Schr\"{o}dinger
equation whereas $u_{n}(r)$ is a moderating function corresponding to the
perturbing potential. Following the prescription of Ref. 24, we may rewrite
(2) as

\begin{equation}
\frac{\hbar ^{2}}{2m}\left( \frac{\chi _{n}^{\prime \prime }(r)}{\chi _{n}(r)%
}+\frac{u_{n}^{\prime \prime }(r)}{u_{n}(r)}+2\frac{\chi _{n}^{\prime
}(r)u_{n}^{\prime }(r)}{\chi _{n}(r)u_{n}(r)}\right) =V(r)-E_{n}.
\end{equation}
The logarithmic derivatives of the unperturbed $\chi _{n}(r)$ and perturbed $%
u_{n}(r)$ wave functions are given by

\begin{equation}
W_{n}(r)=-\frac{\hbar }{\sqrt{2m}}\frac{\chi _{n}^{\prime }(r)}{\chi _{n}(r)}%
\text{ \ \ and \ \ }\Delta W_{n}=-\frac{\hbar }{\sqrt{2m}}\frac{%
u_{n}^{\prime }(r)}{u_{n}(r)},
\end{equation}
which leads to

\begin{equation}
\frac{\hbar ^{2}}{2m}\frac{\chi _{n}^{\prime \prime }(r)}{\chi _{n}(r)}%
=W_{n}^{2}(r)-\frac{\hbar }{\sqrt{2m}}W_{n}^{^{\prime }}(r)=\left[ V_{0}(r)+%
\frac{\hbar ^{2}}{2m}\frac{\ell (\ell +1)}{r^{2}}\right] -\varepsilon _{n},
\end{equation}
where $\varepsilon _{n}$ is the eigenvalue for the exactly solvable
potential of interest, and

\begin{equation}
\frac{\hbar ^{2}}{2m}\left( \frac{u_{n}^{\prime \prime }(r)}{u_{n}(r)}+2%
\frac{\chi _{n}^{\prime }(r)u_{n}^{\prime }(r)}{\chi _{n}(r)u_{n}(r)}\right)
=\Delta W_{n}^{2}(r)-\frac{\hbar }{\sqrt{2m}}\Delta W_{n}^{\prime
}(r)+2W_{n}(r)\Delta W_{n}(r)=\Delta V(r)-\Delta \varepsilon _{n},
\end{equation}
in which $\Delta \varepsilon _{n}=E_{n}^{(1)}+E_{n}^{(2)}+\cdots $ is the
correction term to the energy due to $\Delta V(r)$ and $E_{n}=\varepsilon
_{n}+\Delta \varepsilon _{n}.$ If Eq. (7), which is the most significant
piece of the present formalism, can be solved analytically as in (6), then
the whole problem, in Eq. (2) reduces to the following form

\begin{equation}
\left[ W_{n}(r)+\Delta W_{n}(r)\right] ^{2}-\frac{\hbar }{\sqrt{2m}}%
(W_{n}(r)+\Delta W_{n}(r))^{\prime }=V(r)-E_{n},
\end{equation}
which is a well known treatment within the frame of supersymmetric quantum
theory (SSQT) [26]. Thus, if the whole spectrum and corresponding
eigenfunctions of the unperturbed interaction potential are known, then one
can easily calculate the required superpotential $W_{n}(r)$ for any state of
interest leading to direct computation of related corrections to the
unperturbed energy and wave function.

For the perturbation technique, we can split the given potential in Eq.(2)
into two parts. The main part corresponds to a shape invariant potential,
Eq. (6), for which the superpotential is known analytically and the
remaining part is treated as a perturbation, Eq. (7). Therefore, it is
obvious that ECSC potential can be treated using this prescription. In this
case, the zeroth-order term corresponds to the Coulomb potential while
higher-order terms consitute the perturbation. However, the perturbation
term in its present form cannot be solved exactly through Eq. (7). Thus, one
should expand the functions related to the perturbation in terms of the
perturbation parameter $\lambda $,

\begin{equation}
\Delta V(r;\lambda )=\sum_{i=1}^{\infty }\lambda _{i}V_{i}(r),\text{ \ \ }%
\Delta W_{n}(r;\lambda )=\sum_{i=1}^{\infty }\lambda _{i}W_{n}^{(i)}(r),%
\text{ \ }E_{n}^{(i)}(\lambda )=\sum_{i=1}^{\infty }\lambda _{i}E_{n}^{(i)},
\end{equation}
where $i$ denotes the perturbation order. Substitution of the above
expansions into Eq. (7) and equating terms with the same power of $\lambda $%
\ on both sides up to $O(\lambda ^{3})$ gives

\begin{equation}
2W_{n}(r)W_{n}^{(1)}(r)-\frac{\hbar }{\sqrt{2m}}\frac{dW_{n}^{(1)}(r)}{dr}%
=V_{1}(r)-E_{n}^{(1)},
\end{equation}

\begin{equation}
W_{n}^{(1)2}(r)+2W_{n}(r)W_{n}^{(2)}(r)-\frac{\hbar }{\sqrt{2m}}\frac{%
dW_{n}^{(2)}(r)}{dr}=V_{2}(r)-E_{n}^{(2)},
\end{equation}

\begin{equation}
2\left[ W_{n}(r)W_{n}^{(3)}(r)+W_{n}^{(1)}(r)W_{n}^{(2)}(r)\right] -\frac{%
\hbar }{\sqrt{2m}}\frac{dW_{n}^{(3)}(r)}{dr}=V_{3}(r)-E_{n}^{(3)}.
\end{equation}
Hence, unlike the other perturbation theories, Eq. (7) and its expansion,
Eqs. (10-12), give a flexibility for the easy calculations of the
perturbative corrections to energy and wave functions for the $nth$ state of
interest through an appropriately chosen perturbed superpotential.

\section{Application to the ECSC Potential}

\label{A}Considering the recent interest in various power-law potentials in
the literature, we work through the article within the frame of low
screening parameter. In this case, the ECSC potential can be expanded in
power series of the screening parameter $\delta $ as [10]

\begin{equation}
V(r)=-\left( \frac{A}{r}\right) \exp (-\delta r)\cos (\delta r)=-\left(
\frac{A}{r}\right) \sum_{i=0}^{\infty }V_{i}(\delta r)^{i},
\end{equation}
where the perturbation coefficients $V_{i}$ are given by

\begin{equation}
V_{1}=-1,\text{ }V_{2}=0,\text{ }V_{3}=1/3,\text{ }V_{4}=-1/6,\text{ }%
V_{5}=1/30,\cdots .
\end{equation}
We now apply this approximation method to the ECSC potential with the
angular momentum barrier

\begin{equation}
V(r)=-\left( \frac{A}{r}\right) e^{-\delta r}\cos (\delta r)+\frac{\ell
(\ell +1)\hbar ^{2}}{2mr^{2}}=\left[ V_{0}(r)+\frac{\ell (\ell +1)\hbar ^{2}%
}{2mr^{2}}\right] +\Delta V(r),
\end{equation}
where the first piece is the shape invariant zeroth-order which is an
exactly solvable piece corresponding to the unperturbed Coulomb potential
with $V_{0}(r)=-A/r$ while $\Delta V(r)=A\delta -(A\delta
^{3}/3)r^{2}+(A\delta ^{4}/6)r^{3}-(A\delta ^{5}/30)r^{4}+\cdots $ is the
perturbation term. The literature is rich with examples of particular
solutions for such power-law potentials employed in different fields of
physics, for recent applications see Refs. [27,28]. At this stage one may
wonder why the series expansion is truncated at a lower order. This can be
understood as follows. It is widely appreciated that convergence is not an
important or even desirable property for series approximations in physical
problems. Specifically, a slowly convergent approximation which requires
many terms to achieve reasonable accuracy is much less valuable than the
divergent series which gives accurate answers in a few terms. This is
clearly the case for the ECSC problem [29]. However, it is worthwhile to
note that the main contributions come from the first three terms. Thereby,
the present calculations are performed up to the second-order involving only
these additional potential terms, which suprisingly provide highly accurate
results for small screening parameter $\delta .$

\subsection{Ground State Calculations $\left( n=0\right) $}

In the light of Eq. (6), the zeroth-order calculations leading to exact
solutions can be carried out readily by setting the ground-state
superpotential and the unperturbed exact energy as

\begin{equation}
W_{n=0}\left( r\right) =-\frac{\hbar }{\sqrt{2m}}\ \frac{\ell +1}{r}+\sqrt{%
\frac{m}{2}}\frac{A}{(\ell +1)\hbar },\text{ \ \ }E_{n}^{(0)}=-\frac{mA^{2}}{%
2\hbar ^{2}(n+\ell +1)^{2}},\text{ \ \ \ }n=0,1,2,....
\end{equation}
and from the literature, the corresponding normalized Coulomb bound-state
wave function [30]

\begin{equation}
\chi _{n}(r)=N_{n,l}^{(C)}r^{\ell +1}\exp \left[ -\beta r\right] \times
L_{n}^{2\ell +1}\left[ 2\beta r\right] ,
\end{equation}
in which $N_{n,l}^{(C)}=\left[ \frac{2mA}{\left( n+\ell +1\right) \hbar ^{2}}%
\right] ^{\ell +1}\frac{1}{(n+\ell +1)}\frac{1}{\sqrt{\frac{\hbar ^{2}}{mAn!}%
(n+2\ell +1)!}}$ is a normalized constant,\ \ $\beta =\frac{mA}{\left(
n+\ell +1\right) \hbar ^{2}}$ and $L_{n}^{k}\left( x\right)
=\sum_{m=0}^{n}(-1)^{m}\frac{(n+k)!}{\left( n-m\right) !(m+k)!m!}x^{m}$ is
an associate Laguarre polynomial function [31].

For the calculation of corrections to the zeroth-order energy and
wavefunction, one needs to consider the expressions leading to the first-
and second-order perturbation given by Eqs. (10--12). Multiplication of each
term in these equations by $\chi _{n}^{2}(r)$, and bearing in mind the
superpotentials given in Eq. (5), one can obtain the straightforward
expressions for the first-order correction to the energy and its
superpotential:
\begin{equation}
E_{n}^{(1)}=\int_{-\infty }^{\infty }\chi _{n}^{2}(r)\left( -\frac{A\delta
^{3}}{3}r^{2}\right) dr,\text{ }W_{n}^{(1)}\left( r\right) =\frac{\sqrt{2m}}{%
\hbar }\frac{1}{^{X_{n}^{2}(r)}}\int^{r}\chi _{n}^{2}(x)\left[ E_{n}^{(1)}+%
\frac{A\delta ^{3}}{3}x^{2}\right] dx,\
\end{equation}
and also for the second-order correction and its superpotential:

\[
E_{n}^{(2)}=\int_{-\infty }^{\infty }\chi _{n}^{2}(r)\left[ \frac{A\delta
^{4}}{6}r^{3}-W_{n}^{(1)2}\left( r\right) \right] dr,\text{ }
\]
\begin{equation}
W_{n}^{(2)}\left( r\right) =\frac{\sqrt{2m}}{\hbar }\frac{1}{^{X_{n}^{2}(r)}}%
\int^{r}\chi _{n}^{2}(x)\left[ E_{n}^{(2)}+W_{n}^{(1)2}(x)-\frac{A\delta ^{4}%
}{6}x^{3}\right] dx\ ,
\end{equation}
for any state of interest. The above expressions calculate $W_{n}^{(1)}(r)$
and $W_{n}^{(2)}(r)$\ explicitly from the energy corrections $E_{n}^{(1)}$
and $E_{n}^{(2)}$ respectively, which are in turn used to calculate the
moderating wave function $u_{n}(r).$

Thus, through the use of Eqs. (18) and (19), after some lengthy and tedious
integrals, we find the zeeroth order energy shift and their moderating
superpotentials as

\[
E_{0}^{(1)}\ =-\frac{\hbar ^{4}\left( \ell +1\right) ^{2}\left( \ell
+2\right) \left( 2\ell +3\right) }{6Am^{2}}\delta ^{3},
\]

\begin{eqnarray*}
E_{0}^{(2)} &=&\frac{\hbar ^{6}\left( \ell +1\right) ^{3}\left( \ell
+2\right) \left( 2\ell +3\right) \left( 2\ell +5\right) }{24A^{2}m^{3}}%
\delta ^{4} \\
&&-\frac{\hbar ^{10}\left( \ell +1\right) ^{6}\left( \ell +2\right) \left(
2\ell +3\right) \left( 8\ell ^{2}+37\ell +43\right) }{72A^{4}m^{5}}\delta
^{6},
\end{eqnarray*}

\[
W_{0}^{(1)}(r)=-\frac{\hbar \left( \ell +1\right) \delta ^{3}r}{3\sqrt{2m}}%
\left\{ r+\frac{\hbar ^{2}\left( \ell +1\right) \left( \ell +2\right) }{Am}%
\right\} ,
\]
\begin{equation}
W_{0}^{(2)}(r)=-\frac{\hbar \delta ^{4}cr}{2\sqrt{2m}}\left\{ \delta
^{2}r^{3}+ar^{2}+b\left[ r+\frac{\hbar ^{2}(\ell +1)(\ell +2)}{Am}\right]
\right\} -\frac{\hbar \left( \ell +1\right) }{\sqrt{2m}A}E_{0}^{(2)},
\end{equation}
in which

\begin{eqnarray}
a &=&\frac{\hbar ^{2}(\ell +1)(3\ell +7)\delta ^{2}}{Am}-\frac{3Am}{\hbar
^{2}(\ell +1)^{2}},\text{ \ }b=\left[ \frac{\hbar ^{4}(\ell +1)^{2}(8\ell
^{2}+37\ell +43)\delta ^{2}}{2A^{2}m^{2}}-\frac{3}{2}\frac{(2\ell +5)}{(\ell
+1)}\right] ,\text{ }  \nonumber \\
\text{c} &=&\frac{\hbar ^{2}(\ell +1)^{3}}{9Am}
\end{eqnarray}
Therefore, the analytical expressions for the lowest energy and full radial
wave function of an ECSC potential are then given by

\begin{equation}
E_{n=0,\ell }=E_{n=0,\ell }^{(0)}+A\delta +E_{0}^{(1)}+E_{0}^{(2)}+\cdots ,%
\text{ }\psi _{n=0,\ell }(r)\approx \chi _{n=0,\ell }(r)u_{n=0,\ell }(r),
\end{equation}
in which

\begin{equation}
u_{n=0,\ell }(r)\approx \exp \left( -\frac{\sqrt{2m}}{\hbar }\int^{r}\left(
W_{0}^{(1)}\left( x\right) +W_{0}^{(2)}\left( x\right) \right) dx\right) .
\end{equation}
Hence, the explicit form of the full wave function in (22) for the ground
state is

\begin{equation}
\psi _{n=0,\ell }(r)=\left[ \frac{2mA}{(\ell +1)\hbar ^{2}}\right] ^{\ell +1}%
\frac{1}{(\ell +1)^{2}}\sqrt{\frac{Am}{\hbar ^{2}(2\ell +1)!}}r^{\ell
+1}\exp (P(r)),
\end{equation}
with $P(r)=\sum_{i=1}^{5}p_{i}r^{i}$ is a polynomial of fifth order having
the following coefficients:
\begin{equation}
p_{1}=\frac{(\ell +1)}{A}E_{0}^{(2)}-\frac{Am}{(\ell +1)\hbar ^{2}},\text{ \
}p_{2}=\frac{9}{4}\frac{(\ell +2)}{(\ell +1)^{2}}c^{2}d\delta ^{4},\text{ \ }%
p_{3}=\frac{1}{6}cd\delta ^{4},\text{ }p_{4}=\frac{1}{8}ac\delta ^{4},\text{
}p_{5}=\frac{1}{10}c\delta ^{6},\text{\ }
\end{equation}
in which $d=b+\frac{6Am}{\hbar ^{2}(\ell +1)^{2}\delta }$ and other
parameters are given in (21).

\subsection{Excited state calculations $(n\geq 1)$}

The calculations procedures lead to a handy recursion relations in the case
of ground states, but becomes extremely cumbersome in the description of
radial excitations when nodes of wavefunctions are taken into account, in
particular during the higher order calculations. Although several attempts
have been made to bypass this difficulty and improve calculations in dealing
with excited states, (cf. e.g. [32], and the references therein) within the
frame of supersymmetric quantum mechanics.

Using Eqs. (5) and (17), the superpotential $W_{n}(r)$ which is related to
the excited states can be readily calculated through Eqs. (18) and (19). So
the first-order corrections in the first excited state $(n=1)$ are

\[
E_{1}^{(1)}=-\frac{\hbar ^{4}\left( \ell +2\right) ^{2}\left( \ell +7\right)
\left( 2\ell +3\right) }{6Am^{2}}\delta ^{3},
\]
\

\begin{equation}
W_{1}^{(1)}(r)\approx -\frac{\hbar \left( \ell +2\right) \delta ^{3}r}{3%
\sqrt{2m}}\left\{ r+\frac{\hbar ^{2}(\ell +2)(\ell +3)}{Am}\right\} .
\end{equation}
Consequently, the use of $W_{1}^{(1)}(r)$ in the preceeding equation in (19)
gives the energy correction in the second-order as\

\begin{eqnarray}
\ E_{1}^{(2)} &\approx &\frac{\hbar ^{6}\left( \ell +2\right) ^{3}\left(
\ell +11\right) \left( 2\ell +3\right) \left( 2\ell +5\right) }{24A^{2}m^{3}}%
\delta ^{4}  \nonumber \\
&&-\frac{\hbar ^{10}\left( \ell +2\right) ^{6}\left( \ell +3\right) \left(
2\ell +3\right) \left( 7\ell ^{2}+101\ell +211\right) }{72A^{4}m^{5}}\delta
^{6}.
\end{eqnarray}
Therefore, the approximated energy value of the ECSC potential corresponding
to the first excited state is\

\begin{equation}
E_{n=1,\ell }=E_{1}^{(0)}+A\delta +E_{1}^{(1)}+E_{1}^{(2)}+\cdots .
\end{equation}

The related radial wavefunction can be expressed in an analytical form in
the light of Eqs (18), (19) and (22), if required. The appromation used in
this work would not affect considerably the sensitivity of the calculations.
On the other hand, it is found analytically that our investigations put
forward an interesting hierarchy between $W_{n}^{(1)}(r)$ terms of different
quantum states in the first order after circumventing the nodal difficulties
elegantly,\ \ \

\begin{equation}
W_{n}^{(1)}(r)=-\frac{\hbar \left( n+\ell +1\right) \delta ^{3}r}{3\sqrt{2m}}%
\left\{ r+\frac{\hbar ^{2}(n+\ell +1)(n+\ell +2)}{Am}\right\} ,
\end{equation}
which, for instance, for the second excited state $\left( n=2\right) $ leads
to the first-order correction

\[
\ E_{2}^{(1)}=-\frac{\hbar ^{4}\left( \ell +3\right) ^{2}\left( \ell
+2\right) \left( 2\ell +23\right) }{6Am^{2}}\delta ^{3},
\]

\begin{equation}
W_{2}^{(1)}(r)=-\frac{\hbar \left( \ell +3\right) \delta ^{3}r}{3\sqrt{2m}}%
\left\{ r+\frac{\hbar ^{2}(\ell +3)(\ell +4)}{Am}\right\} .
\end{equation}
Thus, the use of $W_{2}^{(1)}(r)$ in the preceeding equation (19) gives the
energy correction in the second-order as\

\begin{eqnarray}
\ E_{2}^{(2)} &=&\frac{\hbar ^{6}\left( \ell +2\right) \left( \ell +3\right)
^{2}\left( 2\ell +5\right) \left( 2\ell ^{2}+45\ell +153\right) }{%
24A^{2}m^{3}}\delta ^{4}  \nonumber \\
&&-\frac{\hbar ^{10}\left( \ell +2\right) \left( \ell +3\right) ^{5}(16\ell
^{4}+474\ell ^{3}+3879\ell ^{2}+12118\ell +12873)}{72A^{4}m^{5}}\delta ^{6}.
\end{eqnarray}
Therefore, the approximated energy eigenvalue of the ECSC potential
corresponding to the second excited state is\

\begin{equation}
E_{n=2,\ell }=E_{2}^{(0)}+A\delta +E_{2}^{(1)}+E_{2}^{(2)}+\cdots .
\end{equation}
For the numerical work, some numerical values of the perturbed energies of
the $1s$ and $2s$ states, in the atomic units we take $\hbar =m=A=1,$ for
different values of the screening parameter $\delta $ in the range $0\leq
\delta \leq 0.10$ are displayed in Tables 1 and 2, respectively. The results
are consistent to order $\delta ^{6\text{ }}$with earlier results obtained
by applying different methods in Refs. [9,22,23]. Further, we display the
results for the energy eigenvalues of $2s,$ $2p,$ $3s,$ $3p,$ and $3d$
states in Tables 3 and 4. Our results are then compared with accurate energy
eigenvalues obtained by other authors. Thus, through the comparison of our
results with those of Refs. [9,10,22,23] for large $n$ and $\ell -$ values
and small screening parameter values yields indeed excellent results.

On the other hand, we take $A=\sqrt{2}$ and $\delta =\sqrt{2}G.$
Cosequently, we compute the binding energies $(-E_{n,\ell })$ of the
lowest-lying states, $1s$ to $3d,$ for various values of $\delta .$ Hence,
the detailed analysis of the results in terms of various domains of
parameters $A$ and $\delta $ of ECSC potential are displayed in Table 5. For
further study of the bound-state energies and normalizations with analytical
perturbation calculation in Table 6. We consider $A=Z=4,$ $8,$ $16,$ 24 in
order to cover the range of low to high atomic numbers. For low strength of $%
A=Z,$ the energy eigenvalues nobtained are in good agreement with the other
methods for low values of the screening parameter $\delta .$ Obviously, when
$\delta $ is small the Coulomb field character prevails and the method has
been adjusted to that. However, the results become gradually worse as $A$
and/or $\delta $ are large.

\section{Concluding Remarks}

\label{CR}We have shown that the bound-state energies of the exponential
cosine screened Coulomb (ECSC) potential for all eigenstates can be
accurately determined within the framework of a new approximation formalism.
Avoiding the disadvantages of the standard non-relativistic perturbation
theories, the present formulae have the same simple form both for ground and
excited states and provide, in principle, the calculation of the
perturbation corrections up to any arbitrary order in analytical or
numerical form.

Additionally, the application of the present technique to ECSC potential is
really of great interest leading to analytical expressions for both energy
eigenvalues and wave functions. Comparing various energy levels with
different works in the literature we find that this treatment is quite
reliable and further analytical calculations with this non-perturbative
scheme would be useful. In particular, the method becomes more reliable as
the potential strength increases.

\acknowledgments S.M. Ikhdair wishes to dedicate this work to his son Musbah
for his love and assistance.

\bigskip

\begin{table}[tbp]
\caption{Comparison of bound energy eigennvalues for $0\leq \protect\delta
\leq 0.1$ for the $1s$ state in atomic units.}
\begin{tabular}{lllll}
$\delta $ & $1/N$ [22] & Dynamical [9] & Shifted $1/N$ [23] & $E_{n,\ell }$
\\
\tableline0.01 & $-0.490$ $001$ & $-0.490$ $001$ $0$ &  & $-0.490$ $000$ $9$
\\
0.02 & $-0.480$ $008$ & $-0.480$ $007$ $8$ & $-0.480$ $007$ $83$ & $-0.480$ $%
007$ $8$ \\
0.03 & $-0.470$ $026$ & $-0.470$ $026$ $0$ &  & $-0.470$ $025$ $9$ \\
0.04 & $-0.460$ $061$ & $-0.460$ $060$ $9$ & $-0.460$ $061$ $01$ & $-0.460$ $%
060$ $8$ \\
0.05 & $-0.450$ $117$ & $-0.450$ $117$ $4$ &  & $-0.450$ $117$ $2$ \\
0.06 & $-0.440$ $200$ & $-0.440$ $200$ $4$ & $-0.440$ $200$ $57$ & $-0.440$ $%
200$ $0$ \\
0.07 & $-0.430$ $313$ &  &  & $-0.430$ $313$ $4$ \\
0.08 & $-0.420$ $461$ & $-0.420$ $463$ $6$ & $-0.420$ $463$ $86$ & $-0.420$ $%
461$ $7$ \\
0.09 & $-0.410$ $647$ &  &  & $-0.410$ $648$ $8$ \\
0.1 & $-0.400$ $875$ & $-0.400$ $883$ $9$ & $-0.400$ $884$ $21$ & $-0.400$ $%
878$ $5$%
\end{tabular}
\end{table}

\begin{table}[tbp]
\caption{Comparison of bound energy eigennvalues for $0\leq \protect\delta
\leq 0.1$ for the $2s$ state in atomic units.}
\begin{tabular}{lllll}
$\delta $ & $1/N$ [22] & Dynamical [9] & Shifted $1/N$ [23] & $E_{n,\ell }$
\\
\tableline0.01 & $-0.115$ $013$ & $-0.115$ $013$ $5$ &  & $-0.115$ $013$ $4$
\\
0.02 & $-0.105$ $103$ & $-0.105$ $103$ $6$ & $-0.105$ $103$ $61$ & $-0.105$ $%
103$ $3$ \\
0.03 & $-0.095$ $334$ & $-0.095$ $336$ $6$ &  & $-0.095$ $334$ 6 \\
0.04 & $-0.085$ $755$ & $-0.085$ $769$ $0$ & $-0.085$ $769$ $59$ & $-0.085$ $%
762$ $1$ \\
0.05 & $-0.076$ $406$ & $-0.076$ $449$ $7$ &  & $-0.076$ $432$ 6 \\
0.06 & $-0.067$ $311$ & $-0.067$ $421$ $7$ & $-0.067$ $426$ $08$ & $-0.067$
390 0 \\
0.07 & $-0.058$ $482$ &  &  & $-0.058$ 680 0 \\
0.08 & $-0.049$ $915$ & $-0.050$ $392$ $2$ & $-0.050$ $408$ $25$ & $-0.050$
357 6 \\
0.09 & $-0.041$ $598$ &  &  & $-0.042$ 494 5 \\
0.1 & $-0.033$ $500$ & $-0.034$ $967$ $7$ & $-0.035$ $004$ $67$ & $-0.035$
188 0
\end{tabular}
\end{table}

\begin{table}[tbp]
\caption{Energy eigenvalues as a function of screening parameter $\protect%
\delta $ for the $2s$ and $2p$ states in atomic units.}
\begin{tabular}{llllllll}
State & $\delta $ & $E[10,10]$ [10] & $E[10,11]$ [10] & Pertur.[10] &
Variational [10] & Shifted [23] & $E_{n,\ell }$ \\
\tableline$2s$ & $0.10$ & $-0.034$ $941$ & $-0.034$ $941$ & $-0.034$ $425$ &
$-0.034$ $935$ & $-0.035$ $004$ $67$ & $-0.035$ 188 0 \\
$2p$ &  & $-0.032$ $469$ & $-0.032$ $469$ & $-0.032$ $042$ &  & $-0.032$ $%
470 $ $15$ & $-0.032$ 673 3 \\
$2s$ & $0.08$ & $-0.050$ $387$ & $-0.050$ $387$ & $-0.050$ $222$ & $-0.050$ $%
384$ & $-0.050$ $408$ $25$ & $-0.050$ 357 6 \\
$2p$ &  & $-0.048$ $997$ & $-0.048$ $997$ &  &  & $-0.048$ $996$ $93$ & $%
-0.048$ $993$ $9$ \\
$2s$ & $0.06$ & $-0.067$ $421$ & $-0.067$ $421$ & $-0.067$ $385$ & $-0.067$ $%
421$ & $-0.067$ $426$ $08$ & $-0.067$ 390 0 \\
$2p$ &  & $-0.066$ $778$ & $-0.066$ $778$ &  &  & $-0.066$ $777$ $29$ & $%
-0.066$ $761$ $1$ \\
$2s$ & $0.04$ & $-0.085$ $769$ & $-0.085$ $769$ & $-0.085$ $767$ & $-0.085$ $%
769$ & $-0.085$ $769$ $59$ & $-0.085$ $762$ 1 \\
$2p$ &  & $-0.085$ $591$ & $-0.085$ $591$ &  &  & $-0.085$ $559$ $13$ & $%
-0.085$ $552$ $0$ \\
$2s$ & $0.02$ & $-0.105$ $104$ & $-0.105$ $104$ & $-0.105$ $104$ & $-0.105$ $%
104$ & $-0.105$ $103$ $61$ & $-0.105$ $103$ $3$ \\
$2p$ &  & $-0.105$ $075$ & $-0.105$ $075$ & $-0.105$ $075$ &  & $-0.105$ $%
074 $ $64$ & $-0.105$ $074$ 4
\end{tabular}
\end{table}

\begin{table}[tbp]
\caption{Energy eigenvalues as a function of screening parameter $\protect%
\delta $ for the $3s,$ $3p$ and $3d$ states in atomic units.}
\begin{tabular}{llllllll}
State & $\delta $ & $E[10,10]$ [10] & $E[10,11]$ [10] & Pertur. [10] &
Variational [10] & Shifted [23] & $E_{n,\ell }$ \\
\tableline$3s$ & $0.06$ & $-0.005$ $461$ & $-0.005$ $462$ & $-0.004$ $538$ &
$-0.005$ $454$ & $-0.005$ $666$ $38$ & $-0.007$ $077$ $8$ \\
$3p$ &  & $-0.004$ $471$ & $-0.004$ $472$ &  &  & $-0.004$ $492$ $33$ & $%
-0.005$ 405 8 \\
$3d$ &  & $-0.002$ $308$ & $-0.002$ $309$ &  &  & $-0.002$ $313$ $56$ & $%
-0.002$ $924$ $0$ \\
$3s$ & $0.05$ & $-0.011$ $576$ & $-0.011$ $576$ &  &  & $-0.011$ $685$ $44$
& $-0.011$ $952$ $3$ \\
$3p$ &  & $-0.010$ $929$ & $-0.010$ $929$ & $-0.010$ $538$ &  & $-0.010$ $%
939 $ $85$ & $-0.011$ $111$ $7$ \\
$3d$ &  & $-0.009$ $555$ & $-0.009$ $555$ & $-0.009$ $292$ &  & $-0.009$ $%
555 $ $42$ & $-0.009$ $694$ $0$ \\
$3s$ & $0.04$ & $-0.018$ $823$ & $-0.018$ $823$ & $-0.018$ $707$ & $-0.018$ $%
822$ & $-0.018$ $867$ $16$ & $-0.018$ $858$ $6$ \\
$3p$ &  & $-0.018$ $453$ & $-0.018$ $453$ &  &  & $-0.018$ $457$ $05$ & $%
-0.018$ $450$ $5$ \\
$3d$ &  & $-0.017$ $682$ & $-0.017$ $682$ &  &  & $-0.017$ $682$ $08$ & $%
-0.017$ $691$ $0$ \\
$3s$ & $0.02$ & $-0.036$ $025$ & $-0.036$ $025$ & $-0.036$ $022$ & $-0.036$ $%
025$ & $-0.036$ $027$ $38$ & $-0.036$ $021$ $3$ \\
$3p$ &  & $-0.035$ $968$ & $-0.035$ $968$ & $-0.035$ $965$ &  & $-0.035$ $%
967 $ $71$ & $-0.035$ $964$ $0$ \\
$3d$ &  & $-0.035$ $851$ & $-0.035$ $851$ & $-0.035$ $849$ &  & $-0.035$ $%
850 $ $66$ & $-0.035$ $849$ $0$%
\end{tabular}
\end{table}

\begin{table}[tbp]
\caption{Energy eigenvalues of the ECSC potential in units of $\hbar =m=1,$ $%
A=2^{1/2}$ \ and $\protect\delta =GA.$}
\begin{tabular}{lllllllllll}
G & State & $-E_{0,0}$ & State & $-E_{1,0}$ & State & $-E_{0,1}$ & State & $%
-E_{1,1}$ & State & $-E_{0,2}$ \\
\tableline$0.002$ & $1s$ & $0.996$ $000$ $0$ & $2s$ & $0.246$ $000$ $2$ & $%
2p $ & $0.246$ $000$ $1$ & $3p$ & $0.107$ $112$ $0$ & $3d$ & $0.107$ $111$ $%
4 $ \\
$0.005$ &  & $0.990$ $000$ $2$ &  & $0.240$ $003$ $4$ &  & $0.240$ $002$ $4$
&  & $0.101$ $125$ $5$ &  & $0.101$ $116$ $0$ \\
$0.010$ &  & $0.980$ $001$ $9$ &  & $0.230$ $026$ $9$ &  & $0.230$ $019$ $3$
&  & $0.091$ $221$ $7$ &  & $0.091$ $147$ $5$ \\
$0.020$ &  & $0.960$ $015$ $6$ &  & $0.210$ $206$ $6$ &  & $0.210$ $148$ $9$
&  & $0.071$ $928$ $1$ &  & $0.071$ $361$ $7$ \\
$0.025$ &  & $0.950$ $030$ $2$ &  & $0.200$ $395$ $3$ &  & $0.200$ $285$ $7$
&  & $0.062$ $648$ $5$ &  & $0.061$ $566$ $5$ \\
$0.050$ &  & $0.900$ $234$ $4$ &  & $0.152$ $865$ $2$ &  & $0.152$ $099$ $1$
&  & $0.022$ $223$ $5$ &  & $0.014$ $137$ $4$%
\end{tabular}
\end{table}

\begin{table}[tbp]
\caption{Energy eigenvalues of the ECSC potential for all states in units of
$\hbar =2m=1,$ and $\protect\delta =0.2$ $fm^{-1}.$}
\begin{tabular}{llllllll}
$A$ & $\ell $ & $n$ & $-E_{n,\ell }$ & $A$ & $\ell $ & $n$ & $-E_{n,\ell }$
\\
\tableline$4$ & $0$ & $0$ & $3.207$ $029$ & $16$ & $0$ & $1$ & $12.825$ $303$
\\
$8$ & $0$ & $0$ & $14.403$ $752$ &  & $0$ & $2$ & $4.023$ $139$ \\
& $1$ & $0$ & $2.433$ $587$ &  & $1$ & $1$ & $4.009$ $505$ \\
$16$ & $0$ & $0$ & $60.801$ $938$ & $24$ & $0$ & $1$ & $31.217$ $455$ \\
& $1$ & $0$ & $12.818$ $287$ &  & $0$ & $2$ & $11.279$ $786$ \\
$24$ & $0$ & $0$ & $139.201$ $31$ &  & $1$ & $1$ & $11.269$ $899$ \\
& $1$ & $0$ & $31.212$ $563$ &  & $1$ & $2$ & $4.412$ $177$ \\
& $2$ & $0$ & $11.249$ $961$ &  & $2$ & $1$ & $4.380$ $887$ \\
&  &  &  &  & $2$ & $2$ & $1.411$ $568$%
\end{tabular}
\end{table}


\bigskip

\bigskip \bigskip

\bigskip

\begin{references}
\bibitem{1}  P. Anderson, Phys. Rev. 86, 694 (1952); R. Kubo, Phys. Rev. 87,
568 (1952).

\bibitem{2}  R. A. Ferrell and D. J. Scalapino, Phys. Rev. A 9, 846 (1974);
A. J. Bray, J. Phys. A 7, 2144 (1974); E. Brezin, J. Phys. A 12, 759 (1979).

\bibitem{3}  V. L. Bonch-Bruevich and V. B. Glasko, Sov. Phys. Dokl. 4, 147
(1959).

\bibitem{4}  N. Takimoto, J. Phys. Soc. Jpn. 14, 1142 (1959).

\bibitem{5}  V. L. Bonch-Bruevich and Sh. M. Kogan, Sov. Phys. Solid State
1,1118 (1960) C. Weisbuch and B. Vinter, Quantum Semiconductor
Heterostructures (Academic Press, New York, 1993); P. Harrison, Quantum
Wells, Wires and Dots (John Wiley and Sons, England, 2000).

\bibitem{6}  E. P. Prokopev, Sov. Phys. Solid State 9, 993 (1967).

\bibitem{7}  R. Dutt and Y. P. Varshni, Z. Phys. D 2, 207 (1986).

\bibitem{8}  R. Dutt, A. Ray and P. P. Ray, Phys. Lett. A 83, 65 (1981); C.
S. Lam and Y. P. Varshni, Phys. Rev. A 4, 1875 (1971); D. Singh and Y. P.
Varshni, Phys. Rev. A 28, 2606 (1983).

\bibitem{9}  H. de Meyer{\it \ et al}., J. Phys. A 18, L 849 (1985).

\bibitem{10}  C. S. Lai, Phys. Rev. A 26, 2245 (1982).

\bibitem{11}  R. Dutt {\it et al}., J. Phys. A 18, 1379 (1985).

\bibitem{12}  C. S. Lam and Y. P. Varshni, Phys. Rev. A 6, 1391 (1972).

\bibitem{13}  V. L. Eletsky, V. S. Popov, and V. M. Weinberg, Phys. Lett. A
84, 235 (1981).

\bibitem{14}  A. D. Dolgov and V. S. Popov, Phys. Lett. B 79, 403 (1978).

\bibitem{15}  Y. Aharanov and C. K. Au, Phys. Rev. Lett. 42, 1582 (1979)

\bibitem{16}  P. P. Ray and A. Ray, Phys. Lett. B 78, 443 (1981).

\bibitem{17}  C. S. Lai, Phys. Rev. A 23, 455 (1981).

\bibitem{18}  J. D. Hirschfelder, J. Chem. Phys. 33, 1462 (1960).

\bibitem{19}  J. Killingbeck, Phys. Lett. A 65, 87 (1978).

\bibitem{20}  M. Grant and C. S. Lai, Phys. Rev. A 20, 718 (1979).

\bibitem{21}  C. S. Lai, Phys. Rev. A 26, 2245 (1982).

\bibitem{22}  R. Sever and C. Tezcan, Phys. Rev. A 35, 2725 (1987).

\bibitem{23}  S. M. Ikhdair and R. Sever, Z. Phys. D 28,1 (1993).

\bibitem{24}  B. G\"{o}n\"{u}l, K. K\"{o}ksal and E. Bak\i r,
[arXiv:quant-ph/0507098]; B. G\"{o}n\"{u}l, Chin. Phys. Lett. 21, 1685
(2004).

\bibitem{25}  B. G\"{o}n\"{u}l, Chin. Phys. Lett. 21, 2330 (2004); B.
G\"{o}n\"{u}l and M. Ko\c{c}ak, Mod. Phys. Lett. A 20, 355 (2005); B.
G\"{o}n\"{u}l, N. \c{C}elik and E. Ol\u{g}ar, Mod. Phys. Lett. A 20, 1683
(2005); B. G\"{o}n\"{u}l and M. Ko\c{c}ak, Chin. Phys. Lett. 20, 2742
(2005); ibid. Mod. Phys. Lett. A 20, 1983 (2005).

\bibitem{26}  F. Cooper, A. Khare and U. P. Sukhatme, Phys. Rep. 251, 267
(1995).

\bibitem{27}  M. Zonjil, J. Math. Chem. 26, 157 (1999).

\bibitem{28}  M. Alberg and L. Wilets, Phys. Lett. A 286, 7 (2001).

\bibitem{29}  D. J. Doren and D. R. Herschbach, Phys. Rev. A 34, 2665 (1986).

\bibitem{30}  H. A. Bethe and E. E. Salpeter, Quantum Mechanics of One- and
Two-Electron Atoms (springer, Berlin, 1957).

\bibitem{31}  L. S. Gradshteyn and I. M. Ryzhik, Tables of Integrals, Series
and Products (Academic, New York, 1965).

\bibitem{32}  C. Lee, Phys. Lett. A 267, 101 (2000).
\end{references}
\end{document}